\documentclass[12pt,epsf]{article}
\usepackage{axodraw}
\usepackage{graphicx}
\usepackage{enumerate}
\def\lsim{\mathrel{\vcenter{\hbox{$<$}\nointerlineskip\hbox{$\sim$}}}}
\newcommand{\be}{\begin{equation}}
\newcommand{\ee}{\end{equation}}
\newcommand{\ba}{\begin{eqnarray}}
\newcommand{\ea}{\end{eqnarray}}

\def\21{$SU(2) \otimes U(1) $}

\def\lsim{\raise0.3ex\hbox{$\;<$\kern-0.75em\raise-1.1ex\hbox{$\sim\;$}}}
\def\gsim{\raise0.3ex\hbox{$\;>$\kern-0.75em\raise-1.1ex\hbox{$\sim\;$}}} 

\newcommand{\mx}{\left[\begin{array}}
\newcommand{\finmx}{\end{array}\right]} 
\newcommand{\mxp}{\left(\begin{array}} 
\newcommand{\finmxp}{\end{array}\right)} 
\def\beq{\begin{equation}}
\def\eeq{\end{equation}}
\def\bea{\begin{eqnarray}}
\def\eea{\end{eqnarray}}

\def\mathbf#1{\hbox{\bf #1}}
\def\textrm#1{\hbox{#1}}

\def\lsim{\raise0.3ex\hbox{$\;<$\kern-0.75em\raise-1.1ex\hbox{$\sim\;$}}}
\def\gsim{\raise0.3ex\hbox{$\;>$\kern-0.75em\raise-1.1ex\hbox{$\sim\;$}}}
\newcommand {\ignore}[1]{}

\begin{document}
\vspace*{-1in}
\renewcommand{\thefootnote}{\fnsymbol{footnote}}
\begin{flushright}
\texttt{
} 
\end{flushright}
\vskip 5pt
\begin{center}
{\Large{\bf Probing the Majorana mass scale of right-handed neutrinos in mSUGRA
}}
\vskip 25pt

{\sf 
F. Deppisch\footnote[1]{E-mail: deppisch@physik.uni-wuerzburg.de}, 
H. P\"as\footnote[2]{E-mail: paes@physik.uni-wuerzburg.de}, 
A. Redelbach\footnote[3]{E-mail: asredelb@physik.uni-wuerzburg.de}, 
R. R\"uckl\footnote[4]{E-mail: rueckl@physik.uni-wuerzburg.de}}
\vskip 10pt
{\it \small Institut f\"ur Theoretische Physik und Astrophysik\\
Universit\"at W\"urzburg\\ D-97074 W\"urzburg, Germany}\\

\vskip 20pt

{\sf 
Y. Shimizu\footnote[5]{E-mail: shimizu@eken.phys.nagoya-u.ac.jp}}
\vskip 10pt
{\it \small Department of Physics \\ Nagoya University\\ Nagoya, 464-8602, Japan}\\

\vskip 20pt

{\bf Abstract}
\end{center}

\begin{quotation}
{\small 
  
We discuss the perspectives of testing the right-handed Majorana mass scale
$M_R$ of the SUSY see-saw model in the mSUGRA framework. 
Lepton-flavor violating low energy  
processes are analyzed in recently proposed 
post-LEP benchmark scenarios, taking into 
account present uncertainties and future developments in the neutrino sector.
Nonobservation of $\mu \rightarrow e \gamma$ in the next-generation 
PSI experiment will provide upper bounds on $M_R$ of the order of $10^{12\div 14}$~GeV, while on the other hand, a positive signal for 
$\tau \rightarrow \mu \gamma$ at SUPERKEKB or the LHC
may determine $M_R$ for a given mSUGRA scenario 
with an accuracy of a factor of 2.
}
\end{quotation}

\vskip 20pt  

\setcounter{footnote}{0}
\renewcommand{\thefootnote}{\arabic{footnote}}

\newpage
\section{Introduction}

With the evidence for neutrino masses and mixing in solar \cite{sno} 
and atmospheric \cite{sk} neutrino experiments, studies of the lepton sector 
have gained importance as a path to physics beyond the Standard Model. 
The most elegant and widely accepted explanation for 
small neutrino masses is provided by the see-saw mechanism \cite{seesaw},
in which a large Majorana mass scale $M_R$ of right-handed  
neutrinos drives the light neutrino masses 
down to or below the sub-eV scale, as required 
by the experimental evidence. A priori, the fundamental scale $M_R$ can be 
 of the order of the GUT scale, and may thus be 
unaccessible for any kind of direct experimental tests.
However, neutrino mixing implies lepton-flavor violation (LFV), 
which is absent in the Standard Model and provides indirect probes of $M_R$.
While lepton-flavor violating processes are suppressed due to the small 
neutrino masses if only right-handed neutrinos are added to the Standard Model
\cite{petcov}, in supersymmetric models new sources of LFV exist. For example, virtual effects of the massive neutrinos affect the
renormalization group equations (RGE) of the slepton mass and the
trilinear coupling matrices, and give rise to non-diagonal terms inducing LFV.

Assuming the experimentally favored large mixing angle (LMA) MSW solution 
of the solar neutrino anomaly, one can expect lepton-flavor violating $\mu$ and $\tau$ decays with branching ratios close to the current experimental bounds 
\cite{Hisano:1996cp}.
Some of the existing bounds will be improved significantly in the near 
future. The current experimental limits (future sensitivities) on low-energy lepton-flavor violating processes involving charged leptons can be summarized as follows: 
\begin{eqnarray}
&&\bullet\; Br(\mu \rightarrow e \gamma)< 1.2\cdot 10^{-11}(10^{-14})
\mbox{ }\cite{Brooks:1999pu,barkov}\nonumber\\
&&\bullet\; Br(\tau \rightarrow e \gamma)< 2.7\cdot 10^{-6}
\mbox{ }\cite{Groom:in}\nonumber\\
&&\bullet\; Br(\tau \rightarrow \mu \gamma)< 1.1\cdot 10^{-6}(10^{-9})
\mbox{ }\cite{Ahmed:1999gh,superkekb}\label{exp_limits}\\
&&\bullet\; Br(\mu^+\rightarrow e^+e^+e^- )<1.0\cdot 10^{-12}
\mbox{ }\cite{Bellgardt:1987du}\nonumber\\
&&\bullet\; R(\mu^- Ti \rightarrow e^- Ti)<6.1\cdot 10^{-13}
(10^{-14})
\mbox{ } \cite{wintz,sindrum}\nonumber
\end{eqnarray}

\noindent
Here, the observable $R$ denotes the cross-section normalized to the total muon capture rate. The MECO experiment aims at a sensitivity for $\mu^{-} Al\rightarrow e^{-} Al$ below $R \approx 10^{-16}$ \cite{meco}.
In the farther future, the PRISM project plans to provide beams of low-energy 
muons with an intensity increased by several orders of magnitude, so that it 
may become possible to reach 
$Br\left(\mu \rightarrow e \gamma 
\right)\approx 10^{-15}$ \cite{ootani}, 
$Br\left(\mu^+\rightarrow e^+e^+e^- \right)\approx 10^{-16}$ \cite{Aysto:2001zs} and $R\left(\mu^- Ti \rightarrow e^- Ti\right)\approx 10^{-18}$ 
\cite{yoshimura}
(see also the review \cite{Kuno:1999jp}). Searches for 
$\tau \rightarrow \mu \gamma$ at the LHC or SUPERKEKB are expected to probe LFV in this channel at the level of $Br\approx 10^{-9}$
\cite{superkekb}.

The above processes in the context of supersymmetric see-saw models have been 
considered in several previous studies (see e.g. 
\cite{Hisano:1996cp,Hisano:1999fj,Casas:2001sr,topdown,bottomup,other,Kageyama:2001tn}).
In \cite{Hisano:1999fj} it has been pointed out that the corresponding branching ratios and cross-sections exhibit a quadratic dependence on the right-handed Majorana neutrino mass scale $M_{R}$. Therefore, the exploration of these processes provides very interesting possibilities to constrain $M_{R}$. 
In the present paper, we sharpen the current knowledge of these constraints by investigating in more detail which information about the
 right-handed Majorana masses can be extracted from measurements of  
the processes (\ref{exp_limits}). It is assumed that the right-handed Majorana masses are degenerate at the scale \(M_R\). We focus on the recently proposed post-LEP mSUGRA benchmark scenarios \cite{Battaglia:2001zp}, and take into account the uncertainties in the neutrino parameters. In addition, we show by how much the sensitivity to \(M_R\) will improve with future more precise neutrino data. Our work updates and extends previous studies in several directions.
Firstly, the mSUGRA scenarios of \cite{Battaglia:2001zp}  have been developed particularly for linear collider studies, but have not yet been applied to lepton-flavor violating processes at low energies. Our study clarifies the model-dependence of the latter for this very relevant set of mSUGRA models.
Secondly, the neutrino input in our analysis is varied in the ranges allowed by present data. The results are compared to expectations for more precise neutrino measurements in the future. 
Thirdly, we consider degenerate as well as hierarchical neutrino spectra, 
and study the impact, a future determination of the
 absolute neutrino mass scale would make.
Finally, following \cite{Carvalho:2001ex} it is demonstrated that the influence of the mSUGRA scenarios in the tests of $M_{R}$ can be reduced by normalizing 
$Br(l_i \rightarrow l_j \gamma)$ to the corresponding SUSY contribution to
the muon anomalous magnetic moment. 

The paper is organized as follows. In section~2, we discuss the supersymmetric 
see-saw mechanism and the renormalization group evolution of the 
neutrino and slepton mass matrices.
In section~3, the rare decays $l_i \rightarrow l_j \gamma$, 
$\mu \rightarrow 3 e$ as well as $\mu$-$e$ conversion 
in nuclei are briefly reviewed, and the most important results for our investigations are displayed. Also the anomalous magnetic moment of the muon and the 
correlation with $l_{i} \rightarrow l_{j} \gamma$ is discussed there.
Section~4 summarizes the input parameters of the mSUGRA benchmark scenarios and the experimental neutrino data used in the analysis. 
The numerical results of our studies are presented in section~5, and 
conclusions are drawn in section~6.

\newpage
\section{Supersymmetric see-saw mechanism}

The supersymmetric see-saw mechanism is described by the term \cite{Casas:2001sr}
\beq
W_\nu = -\frac{1}{2}\nu_R^{cT}M\nu_R^c + \nu_R^{cT}Y_\nu L\cdot H_2
\eeq
in the superpotential, where \(\nu_{Ra}\) $(a=e, \mu,\tau$) are the right-handed neutrino singlet fields, \(L_a\) denote the left-handed lepton doublets and $H_{2}$ is the Higgs 
doublet with hypercharge $+\frac{1}{2}$.
The $3\times 3$ matrix \(M\) is the Majorana mass matrix, while \(Y_\nu\) is the matrix of neutrino Yukawa couplings leading to the Dirac mass matrix \(m_D=Y_\nu \langle H_2^0 \rangle\), \(\langle H_2^0 \rangle = v\sin\beta\) being the $H_{2}$ vacuum expectation value with \(v=174\)~GeV and \(\tan\beta = \frac{\langle H_2^0\rangle}{\langle H_1^0\rangle}\). Light neutrinos can be naturally explained if one assumes that the Majorana scale \(M_R\) of the mass matrix $M$ is much larger than the scale of the Dirac mass matrix $m_{D}$, which is of the order of the electroweak scale. At energies much smaller than \(M_R\) one has an effective superpotential with
\beq
W^{eff}_{\nu} = \frac{1}{2}(Y_\nu L \cdot  H_2 )^T M^{-1}(Y_\nu L \cdot H_2).
\eeq
The corresponding mass term for the left-handed neutrinos \(\nu_{La}\) is then given by
\beq
 -\frac{1}{2}\nu_{L}^T M_\nu \nu_{L} + h.c.,
\eeq
where the mass matrix
\beq\label{eqn:SeeSawFormula}
M_\nu = m_D^T M^{-1} m_D = Y_\nu^T M^{-1} Y_\nu (v \sin\beta )^2
\eeq
is suppressed by the large Majorana scale $M_{R}$.
In the following we work in the basis where the charged lepton Yukawa coupling matrix \(Y_l\) 
\footnote{Therefore, we do not have to
discriminate flavor and mass eigenstates for charged 
leptons, i.e. \(l_{e,\mu,\tau}=l_{1,2,3}=e,\mu,\tau\)}
and the Majorana mass matrix \(M\) of the right-handed neutrinos are  diagonal, which is always possible. The matrix $M_{\nu}$ is diagonalized by the unitary MNS matrix \(U\),
\beq\label{eqn:NeutrinoDiag}
U^T M_\nu U = \textrm{diag}(m_1, m_2, m_3),
\eeq
that relates the neutrino flavor and mass eigenstates:
\beq
\left(\begin{array}{c}\nu_e \\\nu_\mu \\ \nu_\tau \end{array} \right) 
     =U \left(\begin{array}{c}\nu_1 \\\nu_2 \\ \nu_3 \end{array} \right).
\eeq
In general, \(U\) can be written in the form
\beq\label{eqn:CPphases}
U=V\cdot \textrm{diag}(e^{i\phi_1}, e^{i\phi_2},1),
\eeq
where \(\phi_1,\phi_2\) are Majorana phases and \(V\) can be parametrized in the standard CKM form:
\beq
  V=\left( \begin{array}{ccc} c_{13}c_{12}          & c_{13}s_{12}           & s_{13}e^{-i\varphi}    \\
-c_{23}s_{12}-s_{23}s_{13}c_{12}e^{i\varphi} & c_{23}c_{12}-s_{23}s_{13}s_{12}e^{i\varphi} & s_{23}c_{13} \\
s_{23}s_{12}-c_{23}s_{13}c_{12}e^{i\varphi} &-s_{23}c_{12}-c_{23}s_{13}s_{12}e^{i\varphi} & c_{23}c_{13} 
\end{array} \right).
\eeq
The experimental data on neutrino oscillations determine or at least constrain the mixing matrix \(V\) and the differences of the squared mass eigenvalues \(m_i\) at a scale not far from the electroweak scale. We will therefore identify
these two scales in our analysis. Using the results of recent neutrino fits and making some further necessary assumptions on the neutrino spectrum one can reconstruct \(M_\nu(M_Z)\) from (\ref{eqn:NeutrinoDiag}).

\subsection{Renormalization group evolution of the neutrino sector}
In order to calculate the lepton-flavor violating contributions to the slepton mass matrix in a top-down approach from the unification scale \(M_X\approx 2 \cdot 10^{16}\)~GeV to the electroweak scale, we first need to evolve the neutrino mass matrix \(M_\nu(M_Z)\) to \(M_X\). Below \(M_R\), the one-loop RGE in the MSSM is given by \cite{Ellis:1999my}
\beq\label{eqn:diffMnu}
\frac{d}{dt}M_\nu = \frac{1}{16 \pi^2} \left(\left(-6g_2^2-\frac{6}{5}g_1^2 + Tr(6 Y_U^\dag Y_U)\right)M_\nu + \left((Y_l^\dag Y_l)M_\nu + M_\nu (Y_l^\dag Y_l)^T\right)\right)
\eeq
with the U(1) and SU(2) gauge couplings \(g_1\) and \(g_2\), and the Yukawa coupling matrices \(Y_U\) and \(Y_l\) for the charge $\frac{2}{3}$-quarks and charged leptons, respectively. The corresponding evolution equations for $g_{1,2}$, $Y_{U}$ and $Y_{l}$ can be found in \cite{deBoer:1994dg}. The RGE is linear in \(M_\nu\) and can thus be solved analytically \cite{Ellis:1999my}:
\beq
M_\nu(t) = I(t) \cdot M_\nu(0)\cdot I(t), \quad t=\ln \left(\frac{\mu}{M_Z}\right).
\eeq
Since the evolution is dominated by the gauge and third generation Yukawa couplings one obtains, to a good approximation:
\beq
I(t)=I_g I_t\:\textrm{diag}\left(1,1,I_\tau\right)
\eeq
with
\ba
I_g(t)     &=& \exp \left( \frac{1}{16\pi^2}\int_0^t (-3g_2^2-\frac{3}{5}g_1^2)dt'\right) \\
I_t(t)     &=& \exp \left( \frac{1}{16\pi^2}\int_0^t 3|Y_t|^2 dt'\right) \\
I_\tau(t)  &=& \exp \left( \frac{1}{16\pi^2}\int_0^t |Y_\tau|^2dt'\right).
\ea
To calculate these factors, the MSSM RGEs for the gauge and Yukawa couplings are solved in one-loop approximation, neglecting threshold effects.

In order to proceed with the evolution from \(M_R\) to \(M_X\) we use directly the matrix \(Y_\nu\) of the neutrino Yukawa couplings. From (\ref{eqn:SeeSawFormula}) and (\ref{eqn:NeutrinoDiag}) one finds \cite{Casas:2001sr}
\beq\label{eqn:Ynu}
Y_\nu = \frac{1}{v\sin\beta}\textrm{diag}(\sqrt{M_1},\sqrt{M_2},\sqrt{M_3})\cdot R \cdot \textrm{diag}(\sqrt{m_1},\sqrt{m_2},\sqrt{m_3}) \cdot U^\dag,
\eeq
where \(M_i\) are the Majorana masses of the right-handed neutrinos and \(R\) is an unknown orthogonal matrix. 
As we will see, the lepton-flavor violating terms in the slepton mass matrix depend on \(Y_\nu\) only through the combination \(Y_\nu^\dagger Y_\nu\). In this work, we assume the right-handed Majorana masses to be degenerate at \(M_R\) (\(M_1=M_2=M_3=M_R\)) and the matrix \(R\) to be real. Then the product \(Y_\nu^\dagger Y_\nu\) simplifies to
\begin{eqnarray}\label{eqn:yy}
 Y_\nu^\dagger Y_\nu = \frac{M_R}{v^2\sin^2\beta}U \cdot \textrm{diag}(m_1, m_2, m_3) \cdot U^\dagger,
\end{eqnarray}
thus being independent of \(R\). Therefore this class of models is highly predictive and often used for phenomenological studies. In addition, one obtains more conservative upper bounds on \(M_R\) for real \(R\) because a complex matrix \(R\) generically leads to larger values of \(Y_\nu^\dagger Y_\nu\) and thus to larger branching ratios \(Br(l_i\to l_j \gamma)\) as shown in \cite{Casas:2001sr}.
Furthermore, since the Majorana phases \(\phi_1\) and \(\phi_2\) defined in (\ref{eqn:CPphases}) also drop out in (\ref{eqn:yy}), \(U\) can be replaced by \(V\) in (\ref{eqn:Ynu}). Finally, the neutrino masses \(m_i\) and \(V\) are evaluated from \(M_\nu\) at \(M_R\) using (\ref{eqn:NeutrinoDiag}). The resulting matrix \(Y_\nu (M_R)\) is then evolved from \(M_R\) to \(M_X\) using the one-loop RGE \cite{Casas:2001sr}
\beq\label{eqn:diffYnu}
\frac{d}{dt}Y_\nu = \frac{1}{16 \pi^2}Y_\nu\left(\left(-3g_2^2-\frac{3}{5}g_1^2 + Tr(3 Y_U^\dag Y_U +Y_\nu^\dag Y_\nu)\right)\mathbf{1} + Y_l^\dag Y_l + 3Y_\nu^\dag Y_\nu\right),
\eeq
and keeping the product $Y_{\nu}^{\dagger}Y_{\nu}$ on the r.h.s. of (\ref{eqn:diffYnu}) fixed at $M_{R}$. The running of the right-handed mass matrix \(M\) between \(M_R\) and \(M_X\) is negligible, as we have checked numerically.

\subsection{Renormalization group evolution of the slepton sector}
Having evolved the neutrino Yukawa couplings or, more specifically, the 
product $Y_{\nu}^{\dagger} Y_{\nu}$,
to the unification scale \(M_X\), one can now run the slepton mass matrix from \(M_X\) to the electroweak scale assuming the mSUGRA universality conditions at $M_{X}$:
\begin{equation}
m^{2}_{L}=m_{0}^{2}\mathbf{1},\qquad m^{2}_{R}=m_{0}^{2}\mathbf{1},\qquad A=A_{0}Y_{l} \label{mSUGRAcond},
\end{equation}
where $m_{0}$ is the common scalar mass and $A_{0}$ is the common trilinear coupling.
For the present analysis, we adopt the mSUGRA benchmark scenarios proposed recently in \cite{Battaglia:2001zp} for linear collider studies.
The charged slepton \((\textrm{mass})^2\) matrix has the form:
\begin{equation}\label{ch_slepton_mass_mat}
  m_{\tilde l}^2=\left(
    \begin{array}{cc}
        m^2_{\tilde{l}_{L}}    & (m^{2}_{\tilde{l}_{LR}})^\dagger \\
        m^{2}_{\tilde{l}_{LR}} & m^{2}_{\tilde{l}_{R}}
    \end{array}
      \right),
\end{equation}
where  \(m^2_{\tilde{l}_{L}}\), \(m^{2}_{\tilde{l}_{R}}\) and \(m^{2}_{\tilde{l}_{LR}}\) are \(3\times3\) matrices, \(m^2_{\tilde{l}_{L}}\) and \(m^{2}_{\tilde{l}_{R}}\) being hermitian. The matrix elements are given by
\begin{eqnarray}
  (m^2_{\tilde{l}_L})_{ab}     &=& (m_{L}^2)_{ab} + \delta_{ab}\left(m_{l_a}^2 + m_Z^2 \cos(2\beta)\left(-\frac{1}{2}+\sin^2\theta_W \right)\right) \label{mlcharged} \\
  (m^2_{\tilde{l}_{R}})_{ab}     &=& (m_{R}^2)_{ab} + \delta_{ab}(m_{l_a}^2 - m_Z^2 \cos(2\beta)\sin^2\theta_W) \label{mrcharged} \\
 (m^{2}_{\tilde{l}_{LR}})_{ab} &=& A_{ab}v\cos\beta-\delta_{ab}m_{l_a}\mu\tan\beta,
\end{eqnarray}
\(\theta_W\) being the Weinberg angle and \(\mu\) the SUSY Higgs-mixing parameter. After evolution from $M_{X}$ to $M_{Z}$, one has
\begin{eqnarray}
m_{L}^2&=&m_0^2\mathbf{1} + (\delta m_{L}^2)_{\textrm{\tiny MSSM}} + \delta m_{L}^2 \label{left_handed_SSB} \\
m_{R}^2&=&m_0^2\mathbf{1} + (\delta m_{R}^2)_{\textrm{\tiny MSSM}} + \delta m_{R}^2 \label{right_handed_SSB}\\
A&=&A_0 Y_l+\delta A_{\textrm{\tiny MSSM}}+\delta A \label{A_SSB},
\end{eqnarray}
where
$(\delta m^{2}_{L,R})_{\textrm{\tiny MSSM}}$ and $(\delta A)_{\textrm{\tiny MSSM}}$ denote the usual MSSM renormalization-group corrections \cite{deBoer:1994dg} which are flavor-diagonal.
In addition, the presence of 
right-handed neutrinos radiatively induces flavor off-diagonal terms denoted by $\delta m_{L,R}^2$ and $\delta A$ in (\ref{left_handed_SSB}) to (\ref{A_SSB}).
These corrections are taken into account in the approximation \cite{Hisano:1999fj}
\begin{eqnarray}\label{eq:rnrges}
  \delta m_{L}^2 &=& -\frac{1}{8 \pi^2}(3m_0^2+A_0^2)(Y_\nu^\dag Y_\nu) \ln\left(\frac{M_X}{M_R}\right) \\
  \delta m_{R}^2 &=& 0 \label{delta_m_R} \\
  \delta A &=& -\frac{3 A_0}{16\pi^2}(Y_l Y_\nu^\dag Y_\nu) \ln\left(\frac{M_X}{M_R}\right) .
\end{eqnarray}
It is these terms which give rise to lepton-flavor violating processes such as $l_{i}\rightarrow l_{j}\gamma$ and $\mu$-e conversion.

The physical charged slepton masses are then found by diagonalizing (\ref{ch_slepton_mass_mat}) using the \(6\times6\) unitary matrix \(U_{\tilde l}\):
\begin{equation}
  U_{\tilde l}^{\dagger} m_{\tilde l}^2 U_{\tilde l}=\textrm{diag}(m_{\tilde l_1}^2,...,m_{\tilde l_i}^2,...,m_{\tilde l_6}^2).
\end{equation}
Correspondingly, the slepton mass eigenstates are expressed in terms of the gauge eigenstates by
\begin{equation}
  \tilde l_i=(U_{\tilde l}^{*})_{ai}\tilde l_{La} + (U_{\tilde l}^{*})_{(a+3)i} \tilde l_{Ra}, \quad\quad i=1,...,6; \; a=e,\mu,\tau.
\end{equation}
Similarly to (\ref{mlcharged}), the $3\times3$ (mass\()^2\) matrix of the SUSY partners of the left-handed neutrinos is given by 
\begin{equation}
  (m_{\tilde\nu}^2)_{ab}=(m_{L}^2)_{ab}+\frac{1}{2}\delta_{ab}m_Z^2\cos(2\beta),
\end{equation}
where $m_{L}^2$ can be taken from (\ref{left_handed_SSB}). The partners of the right-handed neutrinos are very heavy and can therefore be disregarded. After diagonalization with the unitary $3\times 3$ matrix \(U_{\tilde\nu}\),
\begin{equation}
  U_{\tilde\nu}^\dagger m_{\tilde\nu}^2 U_{\tilde\nu}=\textrm{diag}(m_{\tilde\nu_1}^2,m_{\tilde\nu_2}^2,m_{\tilde\nu_3}^2),
\end{equation}
the mass eigenstates \(\tilde \nu_i\) are related to the gauge eigenstates by
\begin{equation}
 \left(\begin{array}{c} \tilde \nu_e \\ \tilde \nu_\mu \\ \tilde \nu_\tau\end{array}\right) = U_{\tilde\nu}\left(\begin{array}{c} \tilde \nu_1 \\ \tilde \nu_2 \\ \tilde \nu_3\end{array}\right).
\end{equation}

\section{LFV low-energy processes and $g_{\mu}-2$}
\subsection{The radiative decays $l_{i}\rightarrow l_{j}\gamma$}
The effective Lagrangian for $l_{i}^{-}\rightarrow l_{j}^{-}\gamma$ is given by
\cite{Carvalho:2001ex}
\begin{equation}
\mathcal{L}_{eff}=\frac{e}{2}\bar{l}_{j}\sigma_{\alpha \beta}F^{\alpha \beta}\left(A_{L}^{ij}P_{L}+A^{ij}_{R}P_{R}\right)l_{i},\label{Leff}
\end{equation}
where $F^{\alpha\beta}$ is the electromagnetic field strength tensor, $\sigma_{\alpha\beta}=\frac{i}{2}\left[\gamma_{\alpha},\gamma_{\beta}\right]$ and $P_{R,L}=\frac{1}{2}(1\pm \gamma_{5})$ are the helicity projection operators.
The coefficients $A^{ij}_{L,R}$ are determined by the photon penguin diagrams shown in Fig.~1 with charginos/sneutrinos or neutralinos/charged sleptons in the loop.

\begin{figure} 
\begin{center}
\begin{picture}(400,135)(-200,-50)
\ArrowLine(-190,0)(-150,0)
\Text(-170,-10)[]{$l_{i}$}
\DashLine(-150,0)(-80,0){5}
\Text(-135,-10)[]{$\tilde{l}$}
\Photon(-115,0)(-115,-45){3}{5}
\Text(-100,-30)[]{$\gamma$}
\ArrowLine(-80,0)(-40,0)
\Text(-60,-10)[]{$l_{j}$}
\CArc(-115,0)(35,0,180)
\Text(-115,46)[]{$\tilde{\chi}^{0}$}
\ArrowLine(40,0)(80,0)
\DashLine(80,0)(150,0){5}
\ArrowLine(150,0)(190,0)
\Text(60,-10)[]{$l_{i}$}
\Text(115,-10)[]{$\tilde{\nu}$}
\Photon(115,35)(115,74){3}{5}
\Text(130,55)[]{$\gamma$}
\Text(170,-10)[]{$l_{j}$}
\CArc(115,0)(35,-360,-180)
\Text(88,34)[]{$\tilde{\chi}^{-}$}
\end{picture} 
\caption{\label{lfv_lowenergydiagrams} Diagrams for $l_{i}^{-}\rightarrow l_{j}^{-}\gamma$ in the MSSM}
\end{center}
\end{figure}
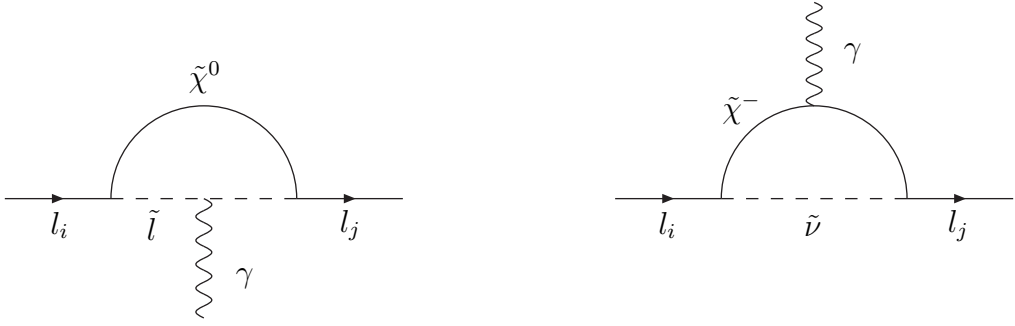 

From (\ref{Leff}) one obtains the following decay rate for $l_{i}^{-}\rightarrow l_{j}^{-}\gamma$ \cite{Hisano:1996cp}:
\begin{equation}
\Gamma\left(l_{i}^{-}\rightarrow l_{j}^{-}\gamma\right)=\frac{\alpha}{4}m^{3}_{l_{i}}\left(\left| A_{L}^{c}+A_{L}^{n}\right|^{2}+\left|A_{R}^{c}+A_{R}^{n}\right|^{2}\right)\label{dec_ij}.
\end{equation}
The superscript \(c\,(n)\) refers to the chargino (neutralino) diagram of Fig.~1, while the flavor indices are omitted.
Because $m_{l_{i}}\gg m_{l_{j}}$ and $m^{2}_{\tilde{l}_{R}}$ is diagonal (see (\ref{mrcharged}), (\ref{right_handed_SSB}) and (\ref{delta_m_R})), one has $A_{R}\gg A_{L}$ \cite{Hisano:1999fj}, \cite{Casas:2001sr}.
The dominant amplitudes in (\ref{dec_ij}) are approximately given by
\begin{eqnarray}
A^{c}_{R}&\simeq& \frac{1}{32\pi^2}\frac{g^{2}_{2}m_{l_{i}}}{\sqrt{2}m_{W}\cos\beta}\sum_{a=1}^{2}\sum_{k=1}^{3}\frac{m_{\tilde{\chi}^{-}_{a}}}{m_{\tilde{\nu}_{k}}^{2}}\left(O_{R}\right)_{a1}\left(O_{L}\right)_{a2}(U_{\tilde{\nu}}^{*})_{jk}(U_{\tilde{\nu}})_{ik}\nonumber\\
& & \times \frac{1}{\left(1-r^{c}_{ak}\right)^{3}}\left(-3+4r^{c}_{ak}-\left(r^{c}_{ak}\right)^{2}-2\ln r^{c}_{ak}\right)\label{domchar}  \\
A^{n}_{R}&\simeq& -\frac{1}{32\pi^{2}}g^{2}_{2}\tan\theta_{W}\sum_{a=1}^{4}\sum_{k=1}^{6}\frac{m_{\tilde{\chi}^{0}_{a}}}{m_{\tilde{l}_{k}}^{2}} \left(O_{N}\right)_{a1}\left(\left(O_{N}\right)_{a2}+\left(O_{N}\right)_{a1}\tan\theta_{W}\right)\nonumber\\
 & &\times (U_{\tilde{l}}^{*})_{jk}(U_{\tilde{l}})_{(i+3)k}\frac{1}{\left(1-r^{n}_{ak}\right)^{3}}\left(1-(r^{n}_{ak})^2+2r^{n}_{ak}\ln r^{n}_{ak}\right)\label{domneut}
\end{eqnarray}
with
\begin{equation}
r^{c}_{ak}=\left(\frac{m_{\tilde{\chi}^{-}_{a}}}{m_{\tilde{\nu}_{k}}}\right)^{2}, \qquad  
r^{n}_{ak}=\left(\frac{m_{\tilde{\chi}^{0}_{a}}}{m_{\tilde{l}_{k}}}\right)^{2},
\end{equation}
the chargino diagonalization matrices $O_{L}$, $O_{R}$ and the neutralino diagonalization matrix $O_{N}$. The mass eigenvalues of the charginos and neutralinos are denoted by $m_{\tilde{\chi}^{-}_{a}}$ and $m_{\tilde{\chi}^{0}_{a}}$, respectively. 
The numerical calculations discussed later are performed with the full expressions for $A_{L}^{c,n}$ and $A_{R}^{c,n}$, which  can be found in \cite{Hisano:1996cp} and \cite{Okada:1999zk}.
\newline
Note that there is no difference between the rates of $l_{i}^{-}\rightarrow l_{j}^{-}\gamma$ and $l_{i}^{+}\rightarrow l_{j}^{+}\gamma$ at the one-loop level and no CP violating observables can be constructed at this level of perturbation theory \cite{Okada:1999zk, Ellis:2001xt}. We therefore do not distinguish between $Br(l_{i}^{-}\rightarrow l_{j}^{-}\gamma)$ and $Br(l_{i}^{+}\rightarrow l_{j}^{+}\gamma)$ in the following.
\newline 
\subsection{$Br\left(\mu\rightarrow 3e\right)$ and 
$R\left(\mu^- N \rightarrow e^- N\right)$}

The processes \(\mu\rightarrow 3e\) and \(\mu^- N \rightarrow e^- N\)
are dominated by photon penguin contributions. As a consequence, one has the following model-independent relations \cite{Hisano:1999fj}: 
\begin{eqnarray}
\frac{Br(\mu\rightarrow 3e)}{Br(\mu\rightarrow e \gamma)} &\approx& \frac{\alpha}{8\pi}\frac{8}{3}\left(\ln\frac{m_{\mu}^{2}}{m_{e}^{2}}-\frac{11}{4}\right) \;\approx\;  7\cdot 10^{-3} \label{mu3erel} \\
\frac{R(\mu^- N \rightarrow e^- N)}{Br(\mu\rightarrow e \gamma)} &\approx& \frac{\Gamma_{\mu}}{\Gamma_{cap}}16\alpha^{4}Z_{eff}^{4}Z|F(q^2)|^{2} \\ 
&\approx& 6\cdot 10^{-3} \qquad \mbox{for Titanium} \label{mueconvrel},
\end{eqnarray}
where $\Gamma_{\mu}$ is the total decay width of the muon, $F(q^{2})$ is the nuclear form factor and $Z$ ($Z_{eff}$) is the electric (effective) charge of the nucleus. From the above and (\ref{exp_limits}) one can see that the present experimental upper limits on $Br\left(\mu\rightarrow 3e\right)$ and $R\left(\mu^- N \rightarrow e^- N\right)$ constrain LFV considerably less than the current limit on $Br\left(\mu\rightarrow e \gamma\right)$.
However, a future measurement of $R$ in the range of $R\left(\mu^- Ti \rightarrow e^- Ti\right)\approx 10^{-18}$ \cite{yoshimura} as mentioned in the introduction would provide a more sensitive test than the corresponding future sensitivity $Br\left(\mu\rightarrow e \gamma\right)\approx 10^{-15}$.

We have examined the relations (\ref{mu3erel}) to (\ref{mueconvrel}) numerically
for the neutrino parameters and mSUGRA scenarios presented in the next section using complete analytic expressions \cite{Hisano:1996cp}, and have found
\begin{eqnarray}
\frac{Br(\mu\rightarrow 3e)}{Br(\mu\rightarrow e \gamma)} &\approx & (6-7)\cdot 10^{-3} \\
\frac{R(\mu^- Ti \rightarrow e^- Ti)}{Br(\mu\rightarrow e \gamma)} &\approx & (5-7)\cdot 10^{-3}
\end{eqnarray}
in good agreement with the above estimates. 

\subsection{Anomalous magnetic moment of the muon} 
The supersymmetric contribution to $\frac{1}{2}(g_{\mu}-2)$ is described by the diagrams of Fig.~1 for $i=j=2$. In this case the effective Lagrangian (\ref{Leff}) yields \cite{Carvalho:2001ex}
\begin{equation} \label{amupeng}
\delta a_{\mu}=\frac{m_{\mu}}{2}\left(A_{R}^{22}+A_{L}^{22}\right)
\end{equation}
and, with (\ref{dec_ij}), the relation
\cite{Carvalho:2001ex}
\begin{equation} \label{amurel}
\frac{Br\left(l_{i} \rightarrow l_{j} \gamma \right)}{\left|\delta a_{\mu}\right|^{2}}\simeq \frac{\alpha}{\Gamma_{i}}\frac{m_{l_{i}}^{3}}{m_{\mu}^{2}}\left|\frac{A_{R}^{ij}}{A_{L}^{22}+A_{R}^{22}}\right|^{2},
\end{equation}
where $\Gamma_{i}$ denotes the total decay width of lepton $l_{i}$ and \(A_R^{ij} \gg A_L^{ij}\) has been used. It will be shown later that this ratio varies less with 
the SUSY parameters and thus
provides a  less model-dependent test of LFV than 
$Br\left(l_{i} \rightarrow l_{j} \gamma \right)$  alone.

\section{Input parameters}
\subsection{mSUGRA benchmark scenarios}
In this paper we focus on the mSUGRA benchmark scenarios proposed in \cite{Battaglia:2001zp}. The theoretical framework of these scenarios is the constrained MSSM with universal soft supersymmetry breaking masses and $R$-parity conservation. Sparticle spectra corresponding to these scenarios are consistent with all experimental and cosmological constraints, in particular with
\begin{itemize}
\item direct sparticle searches
\item $b\rightarrow s \gamma$ 
\item cosmological relic density, with the lightest neutralino as lightest SUSY particle and dark matter candidate 
\item Higgs searches
\end{itemize}

This class of models involves five free parameters: the universal gaugino mass $m_{1/2}$ and the universal scalar mass $m_{0}$ at the GUT scale, the ratio $\tan\beta$ of the Higgs vacuum expectation values, the sign of the Higgs mixing parameter \(\mu\), and the universal trilinear coupling parameter \(A_0\). The values of these parameters for the benchmark scenarios are listed in Tab.~\ref{mSUGRAscen}. $A_{0}$ is set to zero in all scenarios.
Further details can be found in \cite{Battaglia:2001zp}.

\begin{table}[h!]
\begin{center}
\begin{tabular}{|c|c|c|c|c||c|}\hline
Scenario & $m_{1/2}$/GeV  & $m_{0}$/GeV & $\tan\beta$ & sgn$(\mu)$ & $\delta a_{\mu}/10^{-10}$ \\ \hline\hline
A & 600  & 140  &5 & + & 2.8\\ \hline 
B & 250  & 100 &10 & + &28\\ \hline
C & 400  & 90 &10 & + &13\\ \hline
D & 525  & 125 &10 & $-$ &$-7.4$\\ \hline
E & 300  & 1500 & 10& + &1.7\\ \hline
F & 1000  & 3450 & 10& + &0.29\\ \hline
G & 375 & 120 & 20&+  &27\\ \hline
H & 1500 & 419 &20 &+  &1.7\\ \hline
I & 350 & 180 &35 &  +&45\\ \hline
J & 750 & 300 &35 &  +&11\\ \hline
K & 1150 & 1000 &35 & $-$ &$-3.3$\\ \hline
L & 450 & 350 &50 &  +&31\\ \hline
M & 1900 & 1500 &50 & + &2.1\\ \hline
\end{tabular}
\end{center} 
\caption{\label{mSUGRAscen} Input parameters of the mSUGRA benchmark scenarios and the predicted shift $\delta a_{\mu}$ in \(\frac{1}{2}(g_\mu - 2)\) \protect{\cite{Battaglia:2001zp}}.}
\end{table}
\noindent
Also given in Tab.~\ref{mSUGRAscen} is the corresponding shift in the muon anomalous magnetic moment. 
The current $1.6\,\sigma$ discrepancy between the measurement of $\frac{1}{2}(g_{\mu}-2)$ and the Standard Model prediction \cite{Knecht:2001qf} amounts to
\beq\label{eq:anomalousMoment}
\delta a_{\mu}=(25\pm 16 )\cdot 10^{-10}.
\eeq
Scenarios with relatively light sparticle masses below 500 GeV (e.g. B, C, G, L)
are in better agreement with the above value of $\delta a_{\mu}$ 
than scenarios with heavier 
sparticles (e.g. E, F, H, M). 
Moreover, (\ref{eq:anomalousMoment}) favors a positive sign for \(\mu\).
\subsection{Neutrino data}

Solar and atmospheric neutrino experiments provide clear evidence for neutrino oscillations.
 The favored interpretation of the experimental
results on solar neutrinos suggests $\nu_e \rightarrow \nu_{\mu,\tau}$ oscillations driven by 
the mass squared difference $\Delta m^2_{12} = m_2^2 - m_1^2$ in the range of the
LMA solution, while the results on atmospheric neutrinos are interpreted by $\nu_{\mu }\rightarrow \nu_{\tau}$ oscillations driven by $\Delta m^2_{23} = m_3^2 - m_2^2$ in the case of three active neutrinos. 
For the present analysis, we use the global fits in a three-neutrino framework performed in \cite{Gonzalez-Garcia:2001sq}. In \cite{Bahcall:2002hv} it has been pointed out that the inclusion of the SNO result in a two-neutrino analysis of the solar neutrinos implies only minor changes.

We also consider the improvement in our knowledge of the neutrino sector expected from future neutrino experiments, and discuss the consequences of these future accomplishments for the tests of the Majorana scale \(M_R\) considered in this paper. We always assume the present best fit values of the neutrino parameters to remain unchanged. 
The future improvements in the experimental errors of these parameters anticipated for a perspective view are summarized below together with other relevant expectations:

\begin{itemize}

\item {\it $\Delta m_{12}^2$ and $\sin^2 2 \theta_{12}$:}
The long-baseline reactor experiment KAMLAND is designed to test the LMA
MSW solution of the solar neutrino problem. Data taking is expected to start
in 2002 and the solar neutrino parameters will be determined with an accuracy
of $\delta(\Delta m_{12}^2)/\Delta m_{12}^2 = 10 \%$ 
and $\delta(\sin^2 2 \theta_{12})= \pm 0.1$
within three years of measurement
\cite{Barger:2001hy}.

\item {\it $\Delta m_{23}^2$ and $\sin^2 2 \theta_{23}$:}
The atmospheric oscillation parameters will be determined by the long-baseline 
accelerator experiment MINOS with an accuracy of 
$\delta(\Delta m_{23}^2)/\Delta m_{23}^2 = 30 \%$ and $\delta(\sin^2 2 \theta_{23})= \pm 0.1$
\cite{minos}.

\item {\it $\sin^2 2 \theta_{13}$:} 
The CHOOZ reactor experiment
restricts the angle \(\theta_{13}\) to $\sin^2 2 \theta_{13}< 0.1$ \cite{Apollonio:1999ae}. 
The long baseline 
experiment MINOS \cite{minos} can probe the range $\sin^2 2 \theta_{13}\gsim 0.02$-$0.05$.
A future superbeam, a neutrino factory 
\cite{Barger:2000dy}
or the
analysis of the neutrino energy spectra of a future galactic supernova 
\cite{Dighe:2000bi} may 
provide a sensitivity at the level of a few times $10^{-3}$ to 
$10^{-4}$. To explore the potential of
future neutrino studies we take 
\(\delta(\sin^2 2\theta_{13})=3\cdot 10^{-3}\).

\item {\it The neutrino mass spectra:}
The inverse hierarchical spectrum with two heavy and a single light state
is disfavored according to a recent analysis \cite{inv} of the neutrino 
spectrum from supernova SN1987A, unless the mixing angle $\theta_{13}$ is 
large (compare, however, \cite{Barger:2002px}). We therefore restrict 
our analysis to the direct (normal) hierarchy.
LFV rates for inverse hierarchical schemes lie in the intermediate range
between the extreme cases we discuss, $Br(\textrm{degenerate})\ll Br(\textrm{inverse})
<Br(\textrm{hierarchy})$, as pointed out in \cite{Kageyama:2001tn}.

\item {\it The Dirac CP phase $\varphi$:} 
Even at a neutrino factory, one will only be able to 
distinguish \(\varphi = 0\) from $\pi/2$ if $\Delta m^2_{12}>10^{-5}$~eV$^2$.
For this reason we vary $\varphi$ in the full range \(0<\varphi<2\pi\) \cite{Gomez-Cadenas:2001ev}.

\item {\it The neutrino mass scale:}
While neutrino oscillation experiments provide information on the
neutrino mass squared differences $\Delta m^2_{ij}$, the absolute scale of 
the neutrino masses is not known so far. Upper bounds can be obtained from the neutrino 
hot dark matter contribution to the cosmological large scale structure
evolution and the Cosmic Microwave Background, from the interpretation of the extreme energy cosmic rays in the Z-burst model, tritium beta decay experiments and neutrinoless double beta decay experiments \cite{Pas:2001nd}.
A next generation double beta decay experiment like GENIUSI, MAJORANA,
EXO, XMASS or MOON will test the quantity
$m_{ee}= |\sum_i V_{1i}^2 e^{i 2\phi_i} m_i|$ down to $10^{-2}$~eV. Since $V_{13}^2=\sin^2 2 \theta_{13}/4< 0.025$, the contribution of $m_3$ 
drops out and a bound $m_{ee}<10^{-2}$~eV
will imply $m_1< 10^{-2}$~eV$/\cos 2 \theta_{12}$. 
If one further assumes that KAMLAND measures $\sin^2 2 \theta_{12}$
with $\delta(\sin^2 2 \theta_{12})= \pm 0.1$, one
obtains the bound $m_1<3\cdot 10^{-2}$~eV. 
On the other hand, a large mass $m_1$ could be tested by future tritium beta decay
projects. A positive signal at the final sensitivity of the KATRIN experiment 
would imply $m_1=0.3\pm 0.1$~eV \cite{Weinheimer:priv}. Such a value would be compatible with the recent evidence claim for neutrinoless double beta decay \cite{Klapdor-Kleingrothaus:2001ke}.
\end{itemize}
For the present purposes, typical hierarchical and degenerate neutrino mass spectra are parametrized as follows:
\begin{enumerate} [(a)]
\item hierarchical $\nu_{L}$ and degenerate $\nu_{R}$:
\begin{eqnarray}
&&m_{1}\approx 0,~~~m_{2}\approx \sqrt{\Delta m^{2}_{12}},~~~m_{3}\approx\sqrt{\Delta m^{2}_{23}}\\
 &&  M_{1}= M_{2}= M_{3}=M_{R}
\end{eqnarray}
\item quasi-degenerate $\nu_{L}$ and degenerate $\nu_{R}$ \cite{Kageyama:2001tn}:
\begin{eqnarray}
&&m_{1},~~~m_{2}\approx m_1+\frac{1}{2m_1}\Delta m^{2}_{12},~~~m_{3}\approx m_1+\frac{1}{2m_1}\Delta m^{2}_{23}\\
&&M_{1}= M_{2}=M_{3}=M_{R}
\end{eqnarray}
where \(m_1 \gg \sqrt{\Delta m^{2}_{23}} \gg \sqrt{\Delta m^{2}_{12}}.\)
\end{enumerate}
The product of Yukawa couplings $Y_{\nu}^{\dagger}Y_{\nu}$ appearing in the renormalization group corrections to the left-handed slepton mass matrix (\ref{eq:rnrges}) can then be approximated by
\begin{eqnarray}
\mbox{(a) } \left(Y_{\nu}^{\dagger}Y_{\nu}\right)_{ab} &\approx& \frac{M_{R}}{v^{2}\sin^{2}\beta} \sqrt{\Delta m_{23}^2} \left(\sqrt{\frac{\Delta m^{2}_{12}}{\Delta m^{2}_{23}}}V_{a2}V_{b2}^{*}+V_{a3}V_{b3}^{*}\right) \label{llcorrectionhier}\\
\mbox{(b) } \left(Y_{\nu}^{\dagger}Y_{\nu}\right)_{ab} &\approx& \frac{M_{R}}{v^{2}\sin^{2}\beta}\left(m_{1}\delta_{ab}+\frac{\Delta m^{2}_{23}}{2m_1}\left(\frac{\Delta m^{2}_{12}}{\Delta m^{2}_{23}} V_{a2}V_{b2}^{*} + V_{a3}V_{b3}^{*}\right)\right) \label{llcorrectiondeg}.
\end{eqnarray}

In both cases the largest branching ratio for $l_i \rightarrow l_j \gamma$ is expected in the channel $\tau \rightarrow \mu \gamma$ because of 
$|V_{33}V_{23}^*| \approx |V_{32}V_{22}^*|$ and $\Delta m^{2}_{23}\gg \Delta m_{12}^{2}$. The decays $\mu\rightarrow e\gamma$ and $\tau\rightarrow e\gamma$ are suppressed by the smallness of $\Delta m_{12}^{2}$ and $V_{13}$. 
In the case (b), there is an additional suppression by $\sqrt{\Delta m^{2}_{23}}/m_1$ or $\sqrt{\Delta m^{2}_{12}}/m_1$ relative to the case (a). 

In the following analysis, the neutrino parameters are varied in the ranges specified in Tab.~\ref{tab:neutrino_parameters}, characterizing the present knowledge and future prospects.

\begin{figure}[t]
\centering
\includegraphics[clip]{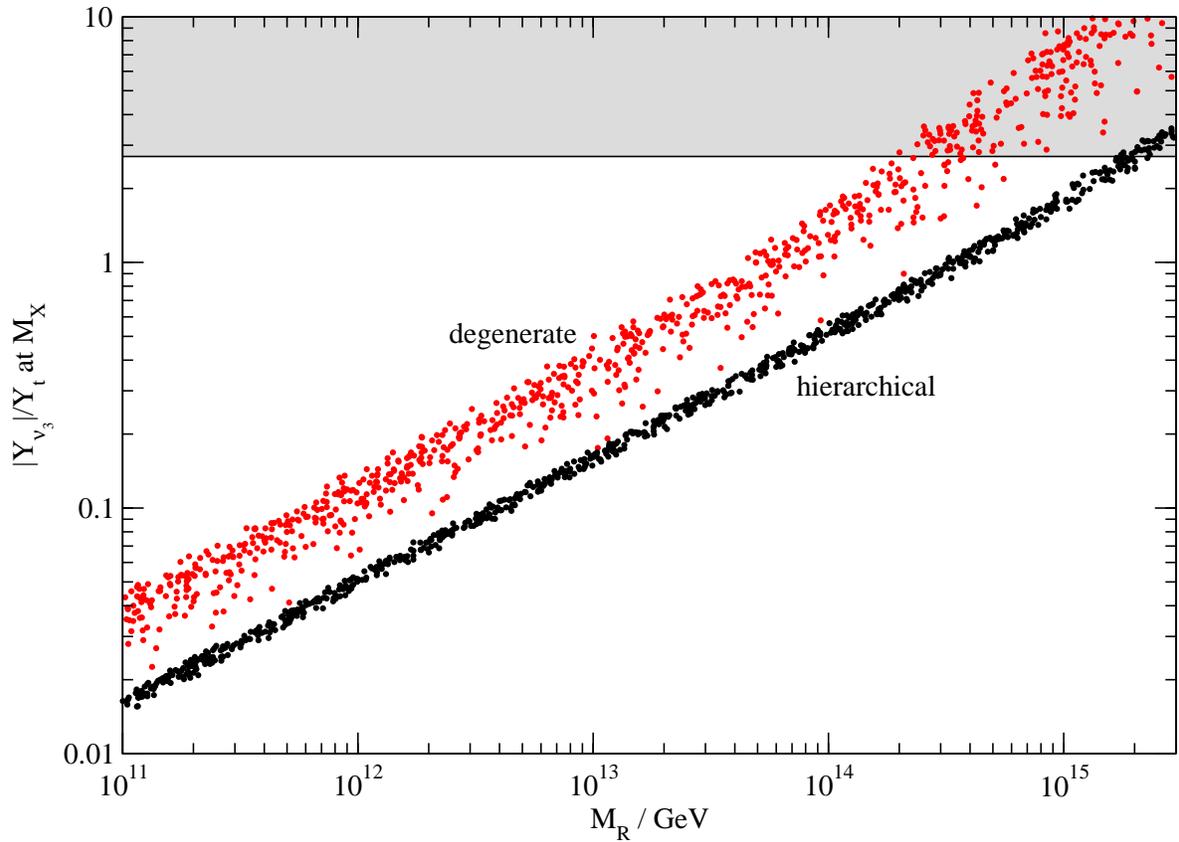}
     \caption{Largest Yukawa coupling \(|Y_{\nu_3}|\) normalized to the top Yukawa coupling \(Y_t\) at \(M_X\) for hierarchical and degenerate neutrino spectra (\(\tan\beta = 30\)). The shaded area is excluded by the constraint \(\frac{|Y_{\nu_3}|^2}{4\pi}<0.3\).}
     \label{fig:yukawa_plots}
\end{figure}

\begin{table}[h]
\begin{center}
\begin{tabular}{|c|c|c|c|}
\hline
parameter                                 & best fit value & present               &  future     \\
\hline
\hline
\(\tan^2\theta_{23}\)                     & 1.40     & \(^{+1.64}_{-1.01}\)  & \(^{+1.37}_{-0.66}\)   \\
\hline
\(\tan^2\theta_{13}\)                     & 0.005    & \(^{+0.050}_{-0.005}\)& \(^{+0.001}_{-0.005}\) \\
\hline
\(\tan^2\theta_{12}\)                     & 0.36     & \(^{+0.65}_{-0.16}\)  & \(^{+0.35}_{-0.16}\)   \\
\hline
\(\Delta m_{12}^2/10^{-5}\textrm{ eV}^2\) & 3.30     & \(^{+66.7}_{-2.3}\)   & \(^{+0.3}_{-0.3}\)     \\
\hline 
\(\Delta m_{23}^2/10^{-3}\textrm{ eV}^2\) & 3.10     & \(^{+3.0}_{-1.7}\)    & \(^{+1.0}_{-1.0}\)     \\
\hline
\hline
\( \varphi/\textrm{rad}\) & \multicolumn{3}{c|}{\(0\textrm{ to } 2\pi\) }  \\
\hline
\( m_1/\textrm{eV} \) & \multicolumn{3}{c|}{\(0\textrm{ to }0.03\) (\(0.3^{+0.11}_{-0.16}\))}         \\
\hline
\end{tabular} 
\end{center}
\caption{90\% CL fits of neutrino parameters characterizing the present and future uncertainties.
The range of the neutrino mass scale \(m_1\) refers to a hierarchical (degenerate) spectrum.
\label{tab:neutrino_parameters}}
\end{table}
Note that all parameters are varied simultaneously. The Majorana mass scale \(M_R\) is treated as a free parameter. This contrasts with other approaches \cite{Hisano:1999fj,Casas:2001sr} where Yukawa coupling unification
\be
|Y_{\nu_3}| = Y_t  \textrm{ at } M_X
\label{yukcouplunif}
\eeq
is assumed, \(|Y_{\nu_3}|^2\) being the largest eigenvalue of \(Y_\nu^\dagger Y_\nu\). Fig.~\ref{fig:yukawa_plots} shows the normalized Yukawa coupling \(|Y_{\nu_3}|/Y_t\) at \(M_X\) as a function of the Majorana mass \(M_R\). One can see that the assumption (\ref{yukcouplunif}) would fix the Majorana mass scale to \(M_R \approx 4\cdot10^{14}\)~GeV for hierarchical neutrinos and to \(M_R \approx 7\cdot10^{13}\)~GeV in the degenerate case. 

Fig.~\ref{fig:yukawa_plots} also shows that for large values of \(M_R\), the Yukawa coupling \(|Y_{\nu_3}|\) eventually gets too strong for perturbation theory to be valid. Therefore, we restrict \(|Y_{\nu_3}|\) to values $\frac{|Y_{\nu_3}|^2}{4\pi}<0.3$, which implies the consistency limits \(M_R < 2\cdot10^{15}\)~GeV in the hierarchical and \(M_R < 3\cdot10^{14}\)~GeV in the degenerate case. 
It should be stressed in this context, that since the negative mass shift \(\delta m_{L}^2\) given in (\ref{eq:rnrges}) is driven by the neutrino Yukawa couplings, the slepton masses decrease with increasing \(M_R\). 
We have checked that in the perturbative region of \(Y_{\nu_3}\) defined above, the slepton masses do not violate existing lower mass bounds, in particular the LEP bound \(m_{\tilde\tau_1} > 81\)~GeV \cite{Groom:in}.

\section{Numerical Results}
The dependence of \(Br(l_i \to l_j \gamma)\) on the right-handed Majorana mass scale \(M_R\) is displayed in Figs.~3, 4 and 5 for those two mSUGRA scenarios listed in Tab.~\ref{mSUGRAscen} which lead to the largest and smallest branching ratios.
The sensitivity on \(M_R\) for all benchmark mSUGRA scenarios defined in Tab.~\ref{mSUGRAscen} is summarized in Tab.~3. The present bounds on $Br(\tau\rightarrow e\gamma)$ and $Br(\tau\rightarrow \mu\gamma)$ set relatively weak constraints on \(M_R\) and are therefore not included in Tab.~3.
For each scenario, the neutrino input is varied in the ranges allowed by present and/or future experiments.
\begin{table}[h!]
\begin{center}
\begin{tabular}{|c|c|c|c|c|}\hline
Scenario & $Br(\mu\rightarrow e \gamma)=1.2\cdot 10^{-11}$ & \multicolumn{2}{|c|}{$Br(\mu\rightarrow e \gamma)=10^{-14}$} & $Br(\tau\rightarrow \mu \gamma)=10^{-9}$ \\ 
&  $M_{R}/10^{14}\mbox{ GeV}$ &  \multicolumn{2}{|c|}{$M_{R}/10^{14}\mbox{ GeV}$} &  $M_{R}/10^{14}\mbox{ GeV}$ \\ 
& present/hier. & future/hier. & future/deg.   & future/hier. \\ \hline\hline
A & [4,20]       &[0.2,4]       &[0.9,3]    &  -      \\ \hline 
B & [0.1,20]    &[0.006,0.4]   &[0.02,2]   &[1,2]    \\ \hline
C & [0.6,20]    &[0.04,0.8]    &[0.2,2]    &[9,11]   \\ \hline
D & [2,20]      &[0.07,2]      &[0.2,2]    &[15,19]  \\ \hline
E & [0.3,20]    &[0.02,0.8]    &[0.1,2]    &[4,5]    \\ \hline
F & [3,20]      &[0.2,2]       &[0.6,2]    &-        \\ \hline
G & [0.2,20]    &[0.01,0.4]    &[0.1,2]    &[2,3]    \\ \hline
H & [4,20]       &[0.3,4]       &[1,3]      &-        \\ \hline
I & [0.04,5]    &[0.003,0.04]  &[0.02,1]   &[0.3,0.6]\\ \hline
J & [0.3,20]    &[0.02,0.8]    &[0.1,2]    &[3,4]    \\ \hline
K & [0.5,20]    &[0.03,0.8]    &[0.2,2]    &[4,7]    \\ \hline
L & [0.04,5]    &[0.003,0.04]  &[0.02,0.6] &[0.2,0.5]\\ \hline
M & [0.6,20]    &[0.06,0.8]    &[0.2,2]    &[6,9]   \\ \hline
\end{tabular}
\caption{\label{intMrhier} Ranges of values for $M_{R}$ for the given branching ratios in the case of hierarchical or degenerate neutrino spectra with present or future uncertainties of neutrino parameters. With '-' we denote that the sensitivity is too low.}
\end{center}
\end{table}

\noindent

Comparing Figs.~3 and 4 with Fig.~5, one can see that  $Br(\mu\rightarrow e \gamma)$ is more strongly affected by the uncertainties in the neutrino parameters than $Br(\tau\rightarrow \mu \gamma)$.
This finding can be understood qualitatively from (\ref{llcorrectionhier}) and (\ref{llcorrectiondeg}), where one sees that $\tau\rightarrow \mu\gamma$ mainly depends on the large angle $\theta_{23}$ while $\mu\rightarrow e \gamma$ involves the small quantities $\theta_{13}$ and $\Delta m^{2}_{12}$. 
The difference in the scatter range of the predictions for $\tau \rightarrow \mu \gamma$ and $\mu \rightarrow e \gamma$ 
thus reflects the different relative error of the quantities \(\theta_{23}, \theta_{13}\) and \(\Delta m_{12}^2\) and also the complete lack of knowledge on $\varphi$ (see Tab. \ref{tab:neutrino_parameters}). 
Furthermore Figs.~3, 4 and 5 and Tab.~\ref{intMrhier} show that the experimental prospects favor the channel \(\mu\to e\gamma\) over \(\tau\to\mu\gamma\) for testing small values of \(M_R\). Larger values of \(M_R\) would be probed more accurately in \(\tau\to\mu\gamma\).
\newline
We also find that for fixed $M_R$ the branching ratios for 
$l_i \rightarrow l_j \gamma$ depend strongly 
on the particular mSUGRA scenario. 
The strongest bounds on $M_{R}$ are obtained in scenario L due to very large $\tan\beta$ and small sparticle masses, whereas scenario H with large gaugino masses yields the weakest bounds.

In summary, we find for hierarchical neutrino spectra that a future measurement of $Br(\tau \rightarrow \mu \gamma)\approx 10^{-9}$ would typically determine $M_{R}$ up to a factor of 2 given the uncertainties in the neutrino parameters.
On the other hand, a measurement of $Br(\mu\rightarrow e\gamma)\approx 10^{-14}$ would determine 
the right-handed scale only up to a factor of 10-100, even if the SUSY parameters would be known. Finally, assuming an exactly massless lightest neutrino (as in previous works), the upper bounds on $M_{R}$ improve by a factor of up to 10. 
For degenerate neutrinos and $M_{R} < 10^{14}$~GeV, $Br(l_i \rightarrow l_j \gamma)$ 
is suppressed 
by roughly two orders of magnitude as compared to the case of hierarchical neutrino spectra,
but exhibit a similar dependence on $M_{R}$. This
can be seen by comparing Figs.~3 and~4. 
However, for $M_{R} > 10^{14}$~GeV, the neutrino Yukawa couplings increase more strongly for degenerate than for hierarchical neutrinos, as illustrated in Fig.~2. Hence for sufficiently large \(M_R\), the branching ratios for hierarchical and degenerate neutrinos become comparable. This behaviour is particularly pronounced for $Br(\tau \rightarrow \mu \gamma)$ as indicated in Fig.~5, because of the enhanced loop contribution from the lightest stau.

\section{Conclusions}
Future experiments searching for lepton-flavor violating rare processes can test the Majorana mass scale $M_R$ of right-handed neutrinos in the see-saw mechanism.
We have systematically and comprehensively studied the sensitivity of $Br(l_{i}\rightarrow l_{j}\gamma)$ on $M_{R}$ in mSUGRA benchmark scenarios designed for future collider studies taking
into account the uncertainties of present and future neutrino 
measurements. We have assumed degenerate Majorana masses for the right-handed neutrinos
and a normal neutrino mass hierarchy, and have considered hierarchical and degenerate 
neutrino spectra. 

For hierarchical neutrinos the measurement of $Br(\mu\rightarrow e\gamma) \approx 10^{-14}$ would probe $M_{R}$ in the range $5\cdot 10^{12}$ ~GeV to $5\cdot 10^{14}$~GeV, depending on the mSUGRA scenario. 
On the other hand, a future measurement of $Br(\tau\rightarrow \mu\gamma)$ at a level of $10^{-9}$ will determine $M_R$ in the range larger than $5 \cdot 10^{13}$~GeV with an accuracy of a factor of 2 for a given scenario. 
In the case of degenerate neutrino masses the upper bound on $M_{R}$ which can be derived from $Br(\mu\rightarrow e\gamma)<10^{-14}$ is $(1 - 3)\cdot 10^{14}$~GeV, independently of the mSUGRA 
scenario. 
Unification of the top Yukawa coupling and the Yukawa coupling of the heaviest neutrino at $M_{X}$ would fix $M_R$ to  
$M_R \approx 4\cdot 10^{14}$~GeV and $M_{R}\approx 7\cdot 10^{13}$~GeV for hierarchical and degenerate neutrinos, respectively. This proposition can thus be tested in the future. 

Planned measurements of $\mu\rightarrow 3e$ are not expected to improve the bounds on $M_R$. On the other hand, a future measurement of $R\left(\mu^- Ti \rightarrow e^- Ti\right)\approx 10^{-18}$ 
is found to be more sensitive to $M_{R}$ by a factor of about 2 than 
$Br\left(\mu\rightarrow e \gamma\right)\approx 10^{-15}$.

The correlation between the SUSY contribution
$\delta a_{\mu}$ to $g_{\mu}-2$ and $Br(l_{i}\rightarrow l_{j}\gamma)$
can be used to reduce the mSUGRA scenario dependence of the above tests. Comparing the ratio \(Br(\tau \rightarrow \mu\gamma)/(\delta a_\mu)^2\) in Fig.~6 with the branching ratio for
\(\tau \rightarrow \mu \gamma \) shown in Fig.~5, one can see that 
for fixed $M_R$ and different mSUGRA scenarios the ratio varies 
by two orders of magnitude less than the value of
$Br(\tau \rightarrow \mu \gamma)$ itself.

\section*{Acknowledgements}   
We thank M. C. Gonzalez-Garcia, J. Hisano, A. Ibarra and
A. Parkhomenko for useful discussions. This work was supported by the Bundesministerium f\"ur Bildung und Forschung (BMBF, Bonn, Germany) under the contract number 05HT1WWA2.

\clearpage
\clearpage
\begin{figure}[t]
\centering
\includegraphics[clip]{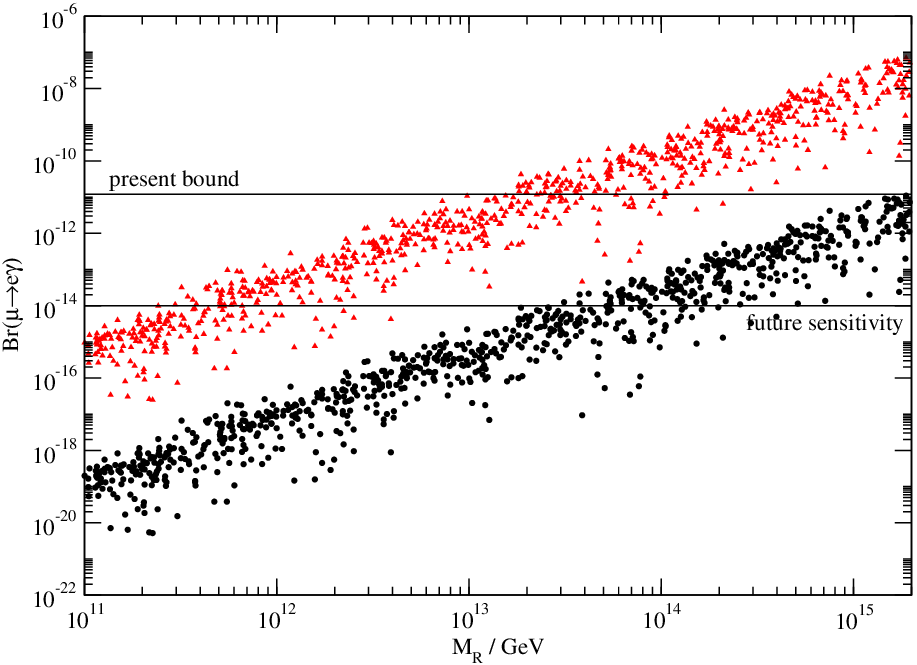}
     \caption{Branching ratio of \(\mu \rightarrow e\gamma\) for hierarchical neutrinos and uncertainties of future neutrino experiments in the mSUGRA scenarios 
leading to the largest (L, upper) and the smallest (H, lower) LFV rates.}
     \label{fig:emu_hier_fut}
\end{figure}
\clearpage
\begin{figure}[t]
\centering
\includegraphics[clip]{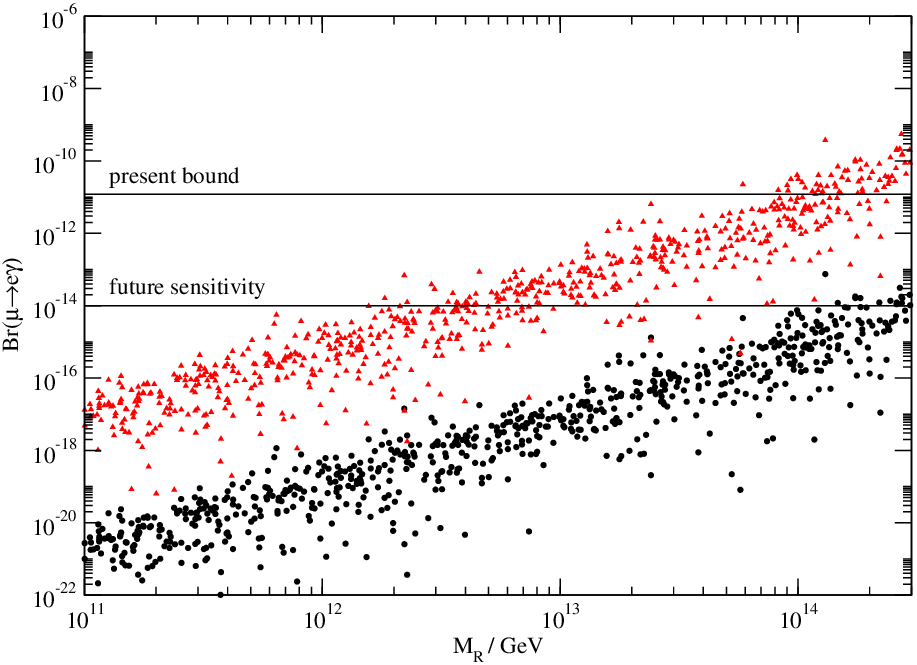}
     \caption{Branching ratio of \(\mu \rightarrow e\gamma\) for degenerate neutrinos  and uncertainties of future neutrino experiments in the mSUGRA scenarios leading to the largest (L, upper) and the smallest (H, lower) LFV rates.}
     \label{fig:emu_deg_fut}
\end{figure}
\clearpage
\begin{figure}[t]
\centering
\includegraphics[clip]{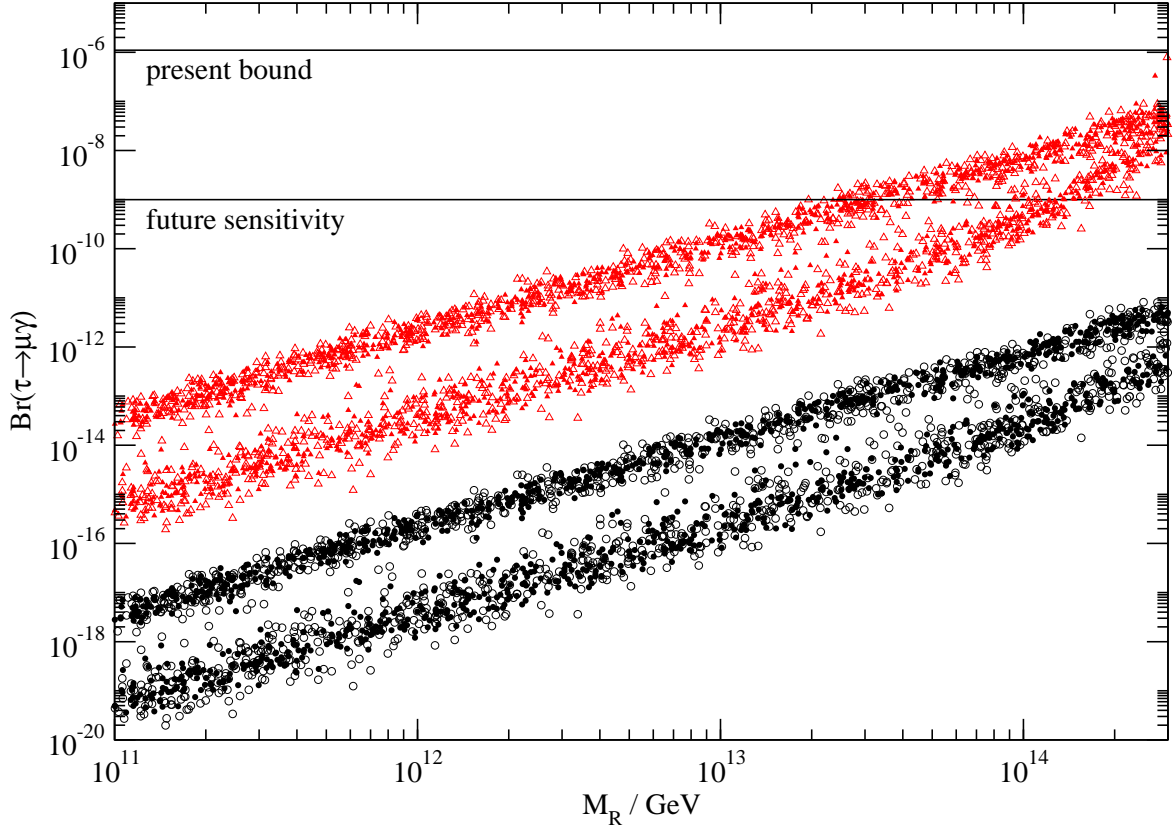}
     \caption{Branching ratio of \(\tau \rightarrow \mu\gamma\) 
for hierarchical (upper) and degenerate (lower) neutrino masses
in the mSUGRA scenarios leading to the largest (L, triangles) and the smallest 
(H, circles) LFV rates.
Open and filled symbols refer to neutrino measurements 
with present and future uncertainties, respectively.}
     \label{fig:mutau_all}
\end{figure}
\clearpage
\begin{figure}[t]
\centering
\includegraphics[clip]{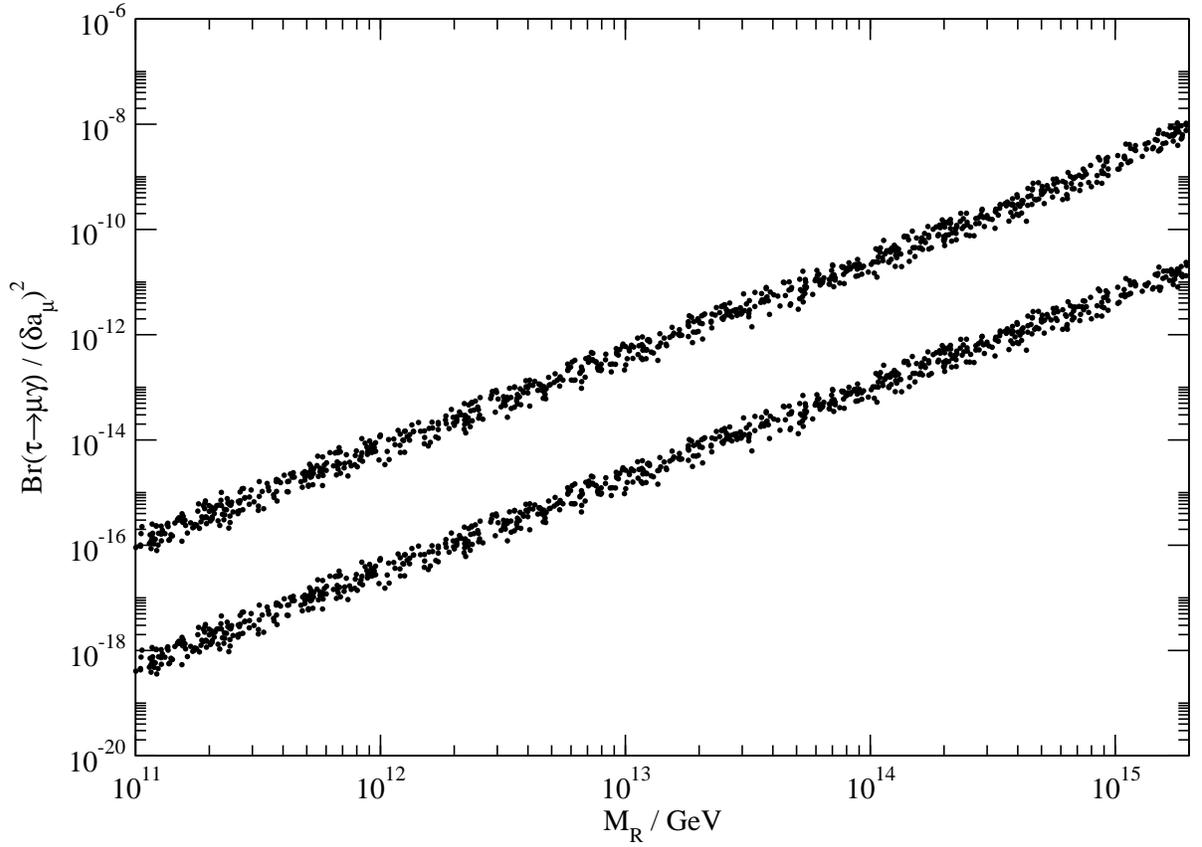}
     \caption{Ratio \(Br(\tau \rightarrow \mu\gamma)/(\delta a_\mu)^2\) 
for hierarchical neutrinos and uncertainties of future neutrino experiments.
Shown are the expectations for the mSUGRA scenarios F (upper) and C (lower) which
embrace the predictions for all other benchmark scenarios of 
Tab.~\ref{mSUGRAscen}.}
     \label{fig:mutau_hier_norm}
\end{figure}

\clearpage


\begin{thebibliography}{99}

\bibitem{sno}
Q.~R.~Ahmad {\it et al.}  [SNO Collaboration],
Phys.\ Rev.\ Lett.\  {\bf 87}, 071301 (2001)
[arXiv:nucl-ex/0106015].

\bibitem{sk} 
Y.~Fukuda {\it et al.}  [Super-Kamiokande Collaboration],
Phys.\ Rev.\ Lett.\  {\bf 81}, 1562 (1998)
[arXiv:hep-ex/9807003].

\bibitem{seesaw}
M. Gell-Mann, P. Ramond and R. Slansky, Proceedings of the Supergravity
Stony Brook Workshop, New York 1979, eds. P. Van Nieuwenhuizen and 
D. Freedman; T. Yanagida, Proceedings of the Workshop on 
Unified Theories and Baryon Number in the Universe, Tsukuba, Japan 1979,
eds. A. Sawada and A. Sugamoto; R. N. Mohapatra and G. Senjanovic, 
Phys. Rev. Lett. {\bf 44}, 912 (1980), {\it erratum} Phys. Rev. {\bf D 23}, 165 
(1993).

\bibitem{petcov}
S.~T.~Petcov,
Sov.\ J.\ Nucl.\ Phys.\  {\bf 25}, 340 (1977)
[Yad.\ Fiz.\  {\bf 25}, 641 (1977\ ERRAT,25,698.1977\ ERRAT,25,1336.1977)];
S.~M.~Bilenkii, S.~T.~Petcov and B.~Pontecorvo,
Phys.\ Lett.\ B {\bf 67}, 309 (1977).

\bibitem{Hisano:1996cp}
J.~Hisano, T.~Moroi, K.~Tobe and M.~Yamaguchi,
Phys.\ Rev.\ D {\bf 53}, 2442 (1996)
[arXiv:hep-ph/9510309].


\bibitem{Brooks:1999pu}
M.~L.~Brooks {\it et al.}  [MEGA Collaboration],
Phys.\ Rev.\ Lett.\  {\bf 83}, 1521 (1999)
[arXiv:hep-ex/9905013].

\bibitem{barkov}
L. M. Barkov {\it et al.,} Research Proposal R-99-05.1 for an experiment at PSI (1999).

\bibitem{Groom:in}
D.~E.~Groom {\it et al.}  [Particle Data Group Collaboration],
Eur.\ Phys.\ J.\ C {\bf 15}, 1 (2000) and 2001 partial update for edition 2002.


\bibitem{Ahmed:1999gh}
S.~Ahmed {\it et al.}  [CLEO Collaboration],
Phys.\ Rev.\ D {\bf 61}, 071101 (2000)
[arXiv:hep-ex/9910060].

\bibitem{superkekb}
L. Serin and R. Stroynowski, ATLAS Internal Note (1997); D. Denegri, private communication;
T. Ohshima, talk at the 3rd Workshop on Neutrino Oscillations and their 
Origin (NOON2001), 2001, ICRR, Univ. of Tokyo, Kashiwa, Japan, to appear in the proceedings.

\bibitem{Bellgardt:1987du}
U.~Bellgardt {\it et al.}  [SINDRUM Collaboration],
Nucl.\ Phys.\ B {\bf 299}, 1 (1988).

\bibitem{wintz}
P.~Wintz, Proceedings of the First International Symposium on Lepton and Baryon Number 
Violation, eds. H.V. Klapdor--Kleingrothaus and I.V. Krivosheina
(Institute of Physics, Bristol, 1998), p. 534.

\bibitem{sindrum}
SINDRUM II collaboration, Research Proposal R-89-06 for an experiment at PSI (1989).

\bibitem{meco}
M. Bachmann {\it et al.} [MECO collaboration], Research Proposal E940 for an 
experiment at BNL (1997).

\bibitem{ootani}
W. Ootani, talk at the 3rd Workshop on Neutrino Oscillations and their 
Origin (NOON2001), 2001, ICRR, Univ. of Tokyo, Kashiwa, Japan, to appear in the proceedings.

\bibitem{Aysto:2001zs}
J.~Aysto {\it et al.}
[arXiv:hep-ph/0109217],
Report of the Stopped Muons Working Group for the ECFA-CERN study on 
Neutrino Factory \& Muon
Storage Rings at CERN.

\bibitem{yoshimura}
K. Yoshimura, talk at the 3rd Workshop on Neutrino Oscillations and their 
Origin (NOON2001), 2001, ICRR, Univ. of Tokyo, Kashiwa, Japan, to appear in the proceedings.

\bibitem{Kuno:1999jp}
Y.~Kuno and Y.~Okada,
Rev.\ Mod.\ Phys.\  {\bf 73}, 151 (2001)
[arXiv:hep-ph/9909265].

\bibitem{Hisano:1999fj}
J.~Hisano and D.~Nomura,
Phys.\ Rev.\ D {\bf 59}, 116005 (1999)
[arXiv:hep-ph/9810479].

\bibitem{Casas:2001sr}
J.~A.~Casas and A.~Ibarra,
Nucl.\ Phys.\ B {\bf 618}, 171 (2001) 
[arXiv:hep-ph/0103065].

\bibitem{topdown}
R.~Barbieri, L.~Hall and A.~Strumia,
Nucl.\ Phys.\ B {\bf 445}, 219 (1995);
G.~K.~Leontaris and N.~D.~Tracas,
Phys.\ Lett.\ B {\bf 431}, 90 (1998);
K.~S.~Babu, B.~Dutta and R.~N.~Mohapatra,
Phys.\ Lett.\ B {\bf 458}, 93 (1999)
[arXiv:hep-ph/9904366];
W.~Buchmuller, D.~Delepine and F.~Vissani,
Phys.\ Lett.\ B {\bf 459}, 171 (1999);
M.~E.~Gomez, G.~K.~Leontaris, S.~Lola and J.~D.~Vergados,
Phys.\ Rev.\ D {\bf 59}, 116009 (1999);
S.~F.~King and M.~Oliveira,
Phys.\ Rev.\ D {\bf 60}, 035003 (1999);
W.~Buchmuller, D.~Delepine and L.~T.~Handoko,
Nucl.\ Phys.\ B {\bf 576}, 445 (2000);
J.~Ellis, M.~E.~Gomez, G.~K.~Leontaris, S.~Lola and D.~V.~Nanopoulos,
Eur.\ Phys.\ J.\ C {\bf 14}, 319 (2000);
J.~L.~Feng, Y.~Nir and Y.~Shadmi,
Phys.\ Rev.\ D {\bf 61}, 113005 (2000);
J.~Sato and K.~Tobe,
Phys.\ Rev.\ D {\bf 63}, 116010 (2001);
D.~F.~Carvalho, M.~E.~Gomez and S.~Khalil,
JHEP {\bf 0107}, 001 (2001);
J.~R.~Ellis, J.~Hisano, M.~Raidal and Y.~Shimizu,
arXiv:hep-ph/0206110.

\bibitem{bottomup}
J.~Sato, K.~Tobe and T.~Yanagida,
Phys.\ Lett.\ B {\bf 498}, 189 (2001);
S.~Davidson and A.~Ibarra,
JHEP {\bf 0109}, 013 (2001).

\bibitem{other}
P.~Ciafaloni, A.~Romanino and A.~Strumia,
Nucl.\ Phys.\ B {\bf 458}, 3 (1996);
J.~Hisano, T.~Moroi, K.~Tobe and M.~Yamaguchi,
Phys.\ Lett.\ B {\bf 391}, 341 (1997);
J.~Hisano, D.~Nomura, Y.~Okada, Y.~Shimizu and M.~Tanaka,
Phys.\ Rev.\ D {\bf 58}, 116010 (1998);
J.~Hisano, D.~Nomura and T.~Yanagida,
Phys.\ Lett.\ B {\bf 437}, 351 (1998);
S.~W.~Baek, N.~G.~Deshpande, X.~G.~He and P.~Ko,
Phys.\ Rev.\ D {\bf 64}, 055006 (2001);
G.~Barenboim, K.~Huitu and M.~Raidal,
Phys.\ Rev.\ D {\bf 63}, 055006 (2001);
X.~J.~Bi and Y.~B.~Dai;
S.~Lavignac, I.~Masina and C.~A.~Savoy,
Phys.\ Lett.\ B {\bf 520}, 269 (2001).

\bibitem{Kageyama:2001tn}
A.~Kageyama, S.~Kaneko, N.~Shimoyama and M.~Tanimoto,
Phys.\ Rev.\ D {\bf 65} 096010 (2002)
[arXiv:hep-ph/0112359].

\bibitem{Battaglia:2001zp}
M.~Battaglia {\it et al.},
Eur.\ Phys.\ J.\ C {\bf 22}, 535 (2001)
[arXiv:hep-ph/0106204].

\bibitem{Carvalho:2001ex}
D.~F.~Carvalho, J.~R.~Ellis, M.~E.~Gomez and S.~Lola,
Phys.\ Lett.\ B {\bf 515}, 323 (2001)
[arXiv:hep-ph/0103256].

\bibitem{Ellis:1999my}
J.~R.~Ellis and S.~Lola,
Phys.\ Lett.\ B {\bf 458}, 310 (1999) 
[arXiv:hep-ph/9904279].

\bibitem{deBoer:1994dg}
W.~de Boer,
Prog.\ Part.\ Nucl.\ Phys.\  {\bf 33}, 201 (1994)
[arXiv:hep-ph/9402266].

\bibitem{Ellis:1999uq}
J.~R.~Ellis, M.~E.~Gomez, G.~K.~Leontaris, S.~Lola and D.~V.~Nanopoulos,
Eur.\ Phys.\ J.\ C {\bf 14}, 319 (2000) 
[arXiv:hep-ph/9911459].

\bibitem{Okada:1999zk}
Y.~Okada, K.~I.~Okumura and Y.~Shimizu,
Phys.\ Rev.\ D {\bf 61}, 094001 (2000) 
[arXiv:hep-ph/9906446].

\bibitem{Ellis:2001xt}
J.~R.~Ellis, J.~Hisano, S.~Lola and M.~Raidal,
Nucl.\ Phys.\ B {\bf 621}, 208 (2002)
[arXiv:hep-ph/0109125].

\bibitem{Knecht:2001qf}
M.~Knecht and A.~Nyffeler,
Phys.\ Rev.\ D {\bf 65}, 073034 (2002)
[arXiv:hep-ph/0111058].

\bibitem{Gonzalez-Garcia:2001sq}
M.~C.~Gonzalez-Garcia, M.~Maltoni, C.~Pena-Garay and J.~W.~Valle,
Phys.\ Rev.\ D {\bf 63}, 033005 (2001)
[arXiv:hep-ph/0009350].

\bibitem{Bahcall:2002hv}
J.~N.~Bahcall, M.~C.~Gonzalez-Garcia and C.~Pena-Garay,
[arXiv:hep-ph/0204314].

\bibitem{Barger:2001hy}
V.~Barger, D.~Marfatia and B.~P.~Wood,
Phys.\ Lett.\ B {\bf 498}, 53 (2001)
[arXiv:hep-ph/0011251].

\bibitem{minos}
MINOS Technical Design Report,\\ 
http://www.hep.anl.gov/ndk/hypertext/minos\_tdr.html.

\bibitem{Apollonio:1999ae}
M.~Apollonio {\it et al.}  [CHOOZ Collaboration],
Phys.\ Lett.\ B {\bf 466}, 415 (1999)
[arXiv:hep-ex/9907037].


\bibitem{Barger:2000dy}
V.~Barger,
talk at Joint U.S. / Japan Workshop on New Initiatives 
in Muon Lepton Flavor Violation and Neutrino Oscillation with 
High Intense
Muon and Neutrino Sources, Honolulu, Hawaii, 2000, 
to appear in the proceedings. 
[arXiv:hep-ph/0102052].

 
\bibitem{Dighe:2000bi}
A.~S.~Dighe and A.~Y.~Smirnov,
Phys.\ Rev.\ D {\bf 62}, 033007 (2000)
[arXiv:hep-ph/9907423].

\bibitem{inv}
H.~Minakata and H.~Nunokawa,
Phys.\ Lett.\ B {\bf 504}, 301 (2001);
%
C.~Lunardini and A.~Y.~Smirnov,
Phys.\ Rev.\ D {\bf 63}, 073009 (2001).

\bibitem{Barger:2002px}
V.~Barger, D.~Marfatia and B.~P.~Wood,
Phys.\ Lett.\ B {\bf 532}, 19 (2002)
[arXiv:hep-ph/0202158].

\bibitem{Gomez-Cadenas:2001ev}
J.~J.~Gomez-Cadenas,
Nucl.\ Phys.\ Proc.\ Suppl.\  {\bf 99B}, 304 (2001)
[arXiv:hep-ph/0105298].

\bibitem{Pas:2001nd}
H.~P\"as and T.~J.~Weiler,
Phys.\ Rev.\ D {\bf 63}, 113015 (2001)
[arXiv:hep-ph/0101091].


\bibitem{Weinheimer:priv}
A.~Osipowicz {\it et al.}  [KATRIN Collaboration],
[arXiv:hep-ex/0109033];
C.~Weinheimer, private communication.

\bibitem{Klapdor-Kleingrothaus:2001ke}
H.~V.~Klapdor-Kleingrothaus, A.~Dietz, H.~L.~Harney and I.~V.~Krivosheina,
Mod.\ Phys.\ Lett.\ A {\bf 16}, 2409 (2001)
[arXiv:hep-ph/0201231].



\end{thebibliography}
\end{document}